\documentstyle[preprint,aps]{revtex}
\begin{document}
\draft
\title{Arbitrary Dimensional Schwarzschild-FRW Black Holes}
\author{Chang Jun Gao$^{}$\thanks{E-mail: gaocj@mail.tsinghua.edu.cn}}
\address{$^{}$Department of Physics and Center for Astrophysics, Tsinghua University, Beijing 100084, China(mailaddress)}

\date{\today}
\maketitle
\begin{abstract}
\hspace*{7.5mm}The metric of arbitrary dimensional Schwarzschild
black hole in the background of Friedman-Robertson-Walker universe
is presented in the cosmic coordinates system. In particular, the
arbitrary dimensional Schwarzschild-de Sitter metric is rewritten
in the Schwarzschild coordinates system and basing on which the
even more generalized higher dimensional Schwarzschild-de Sitter
metric with another extra dimensions is found. The generalized
solution shows that the cosmological constant may roots in the
extra dimensions of space.
\end{abstract}
\section{introduction}
 \hspace*{7.5mm}Black holes are investigated in great depth and
 detail for more than forty years. However, almost all previous studies focused on
isolated black holes. On the other hand, one cannot rule out the
important and more realistic situation in which any black hole is
actually embedded in the background of universe. Therefore black
holes in non-flat backgrounds forms an important topic. \\
\hspace*{7.5mm}As early as in 1933, McVittie [1] found his
celebrated metric for a mass-particle in the expanding universe.
This metric gives us a specific example for a black hole in the
non-flat background. It is just the Schwarzschild black hole which
is embedded in the Friedman-Robertson-Walker universe although
there was no the notion of black hole at that time. To our
knowledge, we have only the McVittie solution that can be used to
describe a black hole which is embedded in our real universe.
Other black hole solutions are in the static backgrounds such as
de Sitter or Einstein spacetime, compared easy and theoretical
cases. They include Schwarzschild-de Sitter (or Einstein),
Reissner-Nordstr$\ddot{o}$m-de Sitter (or Einstein), Kerr-de
Sitter (or Einstein)
and so on [2].\\
\hspace*{7.5mm}Recently, we extended the McVittie's solution into
charged black holes [3]. Tangherlini had the first to generalize
the Schwarzschild solution to higher dimensions [4]. In this
letter, we extended the McVittie's solution from 4 dimensions to
arbitrary dimensions of spacetime. We first deduce the metric of
the Schwarzschild black hole in the FRW
(Friedman-Robertson-Walker) universe. Then we rewrite the
arbitrary dimensional Schwarzschild-de Sitter metric from the
cosmic coordinates system to the Schwarzschild coordinates system.
In order to give an explanation of the origin cosmological
constant $\lambda$, we present a model of the solution of Einstein
equations which shows that $\lambda$ may root
in the extra dimensions of the space.\\
\section{arbitrary dimensional Schwarzschild-FRW black hole}
\hspace*{7.5mm}The metric of $\left(n+3\right)$ dimensional
schwarzschild black hole in the Schwarzschild coordinates system
is given by [5]
\begin{equation}
d{s}^2=-\left(1-\frac{r_0^n}{r^n}\right)
d{t}^2+\left(1-\frac{r_0^n}{r^n}\right)^{-1}d{r}^2+{r}^2d\Omega_{n+1}^2,
\end{equation}
where
 \begin{equation}
r_0^n\equiv \frac{16\pi M}{\left(n+1\right)S_{n+1}}.
\end{equation}
$M$ is the mass of the black hole and $S_{n+1}$ is the area of the
unit $n+1$ dimensional sphere.\\
\hspace*{7.5mm}For our purpose we rewrite the metric Eq.(1)in the
isotropic spherical coordinates. So make variable transformation,
$r\rightarrow x$,
 \begin{equation}
{r}=\frac{x}{2^{\frac{1}{n}}}\left(1+\frac{r_0^n}{2x^n}\right)^{\frac{2}{n}},
\end{equation}
 then we can rewrite Eq.(1) as follows
\begin{equation}
ds^2=-\frac{\left(1-\frac{r_0^n}{2x^n}\right)^2}{\left(1+\frac{r_0^n}{2x^n}\right)^2}dt^2
+\frac{1}{2^{\frac{2}{n}}}{\left(1+\frac{r_0^n}{2x^n}\right)^{\frac{4}{n}}}\left(dx^2+x^2d\Omega_{n+1}^2\right).
\end{equation}
The two constants $\frac{1}{2}$ and $2^{\frac{2}{n}}$ can be
absorbed by $r_0$ and $t$ while without altering the geometry of
the spacetime. So Eq.(4) is identified with
\begin{equation}
ds^2=-\frac{\left(1-\frac{r_0^n}{x^n}\right)^2}{\left(1+\frac{r_0^n}{x^n}\right)^2}dt^2
+{\left(1+\frac{r_0^n}{x^n}\right)^{\frac{4}{n}}}\left(dx^2+x^2d\Omega_{n+1}^2\right).
\end{equation}
\hspace*{7.5mm}On the other hand, the metric for the
$\left(n+3\right)$ dimensional FRW universe is given by
\begin{equation}
 ds^2=-dt^2+\frac{a^2}{\left(1+kx^2/4\right)^2}\left(dx^2+x^2d\Omega_{n+1}^2 \right),
\end{equation}
where $a\equiv a\left(t\right)$ is the scale factor of the
universe which is defined by the homogeneous isotropic matter in
the universe and $k$ gives the curvature of space-time as a whole.
McVittie gave the metric for the Schwarzschild black hole embedded
in the 4 dimensional spactime as follows
\begin{eqnarray}
ds^2=-\frac{\left[\frac{{a^{\frac{1}{2}}}}{{\left(1+kx^2/4\right)^{\frac{1}{2}}}}-\frac{r_0}
{x{a^{\frac{1}{2}}}}\right]^2}{\left[\frac{{a^{\frac{1}{2}}}}{{\left(1+kx^2/4\right)
^{\frac{1}{2}}}}+\frac{r_0}{x{a^{\frac{1}{2}}}}\right]^2}dt^2
+{\left[\frac{{a^{\frac{1}{2}}}}{{\left(1+kx^2/4\right)
^{\frac{1}{2}}}}+\frac{r_0}{x{a^{\frac{1}{2}}}}\right]^{{4}}}\left(dx^2+x^2d\Omega_{2}^2\right).
\end{eqnarray}
When $k=0, a=const$, Eq.(7) restores Eq.(5) ($n=1$). Considering
the metric Eq.(5) we set the metric for the $\left(n+3\right)$
dimensional Schwarzschild-FRW black hole is given by
\begin{equation}
ds^2=-A\left(t,x\right)^2dt^2
+{B\left(t,x\right)^{{\frac{4}{n}}}}\left(dx^2+x^2d\Omega_{n+1}^2\right).
\end{equation}
Consider the Einstein equations
\begin{eqnarray}
 G_{\mu\nu}=8\pi T_{\mu\nu},
 \end{eqnarray}
where $T_{\mu\nu}$ is the energy momentum tensor of the
homogeneous perfect fluid as adopted by McVittie
\begin{equation}
T^{\nu}_{\mu}={diag}\left(\rho, -p, \cdot\cdot\cdot, -p,
\cdot\cdot\cdot\right),
\end{equation}
where $\rho=\rho(t)$ is the energy density and $p=p(t)$ the
pressure of the
perfect fluid.\\
 \hspace*{7.5mm}Then from equation $G_{01}=0$ one obtains
\begin{eqnarray}
A\left(t, x\right)=f\left(t\right)\frac{\dot{B}}{B},
 \end{eqnarray}
where $"\cdot¡±"$ denotes the derivative with respect to $t$.\\
\hspace*{7.5mm}Thinking of Eq.(7) we now take an important step
towards making the deduction very simple. That is to set the
function $B(t, x)$ has the following form
\begin{equation}
B\left(t,x\right)=C\left(x\right)F\left(t\right)+\frac{D\left(x\right)}{F\left(t\right)}.
\end{equation}
\hspace*{7.5mm}Then equations
$G_{1}^1=G^2_2=G^3_3=\cdot\cdot\cdot=8\pi p$ immediately give
\begin{eqnarray}
C\left(x\right)=\frac{1}{\left(1+kx^2/4\right)^{\frac{n}{2}}},\ \
\ \ D\left(x\right)=\frac{r_0^n}{x^n}.
 \end{eqnarray}
In order to recover to the $4$ dimensional case Eq.(7), we have
set the two integration constants in Eq.(13) as $k$ and $r_0^n$.
We also have the form of $F(t)=a(t)^{\frac{n}{2}}$ by inspecting
Eq.(7). Inserting this form and Eq.(12) into Eq.(11), we can
rewrite Eq.(11) as follows
 \begin{equation}
 A=f\left(t\right)\frac{n\dot{a}}{2a}\frac{C\left(x\right)a^{\frac{n}{2}}
 -\frac{D\left(x\right)}{a^{\frac{n}{2}}}}{C\left(x\right)a^{\frac{n}{2}}+\frac{D\left(x\right)}{a^{\frac{n}{2}}}}.
 \end{equation}
Rescale the time variable $t$, $t\rightarrow \tilde{t}$, i.e., set
\begin{equation}
 d\tilde{t}=f\left(t\right)\frac{n\dot{a}}{2a}dt.
 \end{equation}
Then we obtain the $\left(n+3\right)$ dimensional
Schwarzschild-FRW metric
\begin{equation}
ds^2=-\frac{\left[\frac{{a^{\frac{n}{2}}}}{{\left(1+kx^2/4\right)^{\frac{n}{2}}}}-\frac{r_0^n}
{x^n{a^{\frac{n}{2}}}}\right]^2}{\left[\frac{{a^{\frac{n}{2}}}}{{\left(1+kx^2/4\right)
^{\frac{n}{2}}}}+\frac{r_0^n}{x^n{a^{\frac{n}{2}}}}\right]^2}dt^2
+{\left[\frac{{a^{\frac{n}{2}}}}{{\left(1+kx^2/4\right)
^{\frac{n}{2}}}}+\frac{r_0^n}{x^n{a^{\frac{n}{2}}}}\right]^{{\frac{4}{n}}}}\left(dx^2+x^2d\Omega_{n+1}^2\right).
\end{equation}
namely,
\begin{eqnarray}
ds^2&=&-\frac{\left[1-\frac{r_0^n}{a^nx^n}{\left(1+kx^2/4\right)^{\frac{n}{2}}}\right]^2}
{\left[1+\frac{r_0^n}{a^nx^n}{\left(1+kx^2/4\right)^{\frac{n}{2}}}\right]^2}dt^2
+\frac{a^2}{\left(1+kx^2/4\right)^2}{\left[1+\frac{r_0^n}{a^nx^n}
{\left(1+kx^2/4\right)^{\frac{n}{2}}}\right]^{\frac{4}{n}}}\nonumber\\&&\left(dx^2+x^2d\Omega_{n+1}^2\right).
\end{eqnarray}
Here the sign "$\sim$" on $t$ is omitted. When $k=1, a=const$,
Eq.(17) is reduced to the $\left(n+3\right)$ dimensional static
black hole solution, Eq.(5). When $r_0=0$, Eq.(17) is reduced to
the $\left(n+3\right)$ dimensional FRW universe solution, Eq.(6).
When $n=1$, Eq.(17) is just the McVittie solution, Eq.(7). Thus
Eq.(17) describes the $\left(n+3\right)$ dimensional Schwarzschild
black hole solution
which is embedded in the FRW universe.\\
\section{Arbitrary dimensional Schwarzschild-de Sitter black hole}
\hspace*{7.5mm}In this section, we first rewrite the arbitrary
dimensional Schwarzschild-de Sitter metric from the cosmic
coordinates system to the Schwarzschild coordinates system, our
familiar system and then present a new Schwarzschild-de Sitter
solution with another extra dimensions. \\
\hspace*{7.5mm}Set $k=0, a=e^{Ht}$ in Eq.(17), we obtain the
arbitrary dimensional Schwarzschild-de Sitter metric
\begin{equation}
ds^2=-\frac{\left[1-\frac{r_0^n}{a^nx^n}\right]^2}
{\left[1+\frac{r_0^n}{a^nx^n}\right]^2}dt^2
+{a^2}{\left[1+\frac{r_0^n}{a^nx^n}
\right]^{\frac{4}{n}}}\left(dx^2+x^2d\Omega_{n+1}^2\right).
\end{equation}
Make variable transformation below, $x\rightarrow y$,
\begin{equation}
x=2^{-\frac{1}{n}}a^{-1}\left(y^n-2r_0^n+\sqrt{y^{2n}-4r_0^ny^n}\right)^{\frac{1}{n}}.
\end{equation}
Eq.(18) becomes
\begin{equation}
ds^2=-\left(1-\frac{4r_0^n}{y^n}-H^2y^2\right)dt^2+\left(1-\frac{4r_0^n}{y^n}\right)^{-1}dy^2
-2Hy\left(1-\frac{4r_0^n}{y^n}\right)^{-1/2}dtdy+y^2d\Omega_{n+1}^2.
\end{equation}
Eq.(18) has a $dtdy$ term. In order to eliminate this term, we
introduce a new time variable $u$, namely, $t\rightarrow u$
\begin{equation}
t=u-{Hy}{\left(1-\frac{4r_0^n}{y^n}-H^2y^2\right)^{-1}\left(1-\frac{4r_0^n}{y^n}\right)^{-1/2}}.
\end{equation}
Finally in the new coordinates system $(u,y)$, Eq.(18) is reduced
to
\begin{equation}
ds^2=-\left(1-\frac{4r_0^n}{y^n}-H^2y^2\right)du^2+\left(1-\frac{4r_0^n}{y^n}-H^2y^2\right)^{-1}dy^2
+y^2d\Omega_{n+1}^2.
\end{equation}
Absorb the constant $4$ by $r_0$ and rewrite the variables $(t,r)$
instead of $(u,y)$, we obtain the Schwarzschild-de Sitter metric
in the familiar Schwarzschild coordinates system
\begin{equation}
ds^2=-\left(1-\frac{r_0^n}{r^n}-H^2r^2\right)dt^2+\left(1-\frac{r_0^n}{r^n}-H^2r^2\right)^{-1}dr^2
+r^2d\Omega_{n+1}^2.
\end{equation}
When $H=0$, it is just the well known $(n+3)$ dimensional
Schwarzschild metric. When $r_0=0$, it is the $(n+3)$ dimensional
de Sitter metric.  \\
\hspace*{7.5mm}Recently, Gu and Huwang [6] pointed out that the
dark energy perhaps originates from the extra dimensions of the
space. Here we give a model of solution to indicate that the extra
dimensions do can contribute dark energy. We find that the metric
\begin{equation}
ds^2=-\left(1-\frac{r_0^n}{r^n}-\frac{m}{n+2}\frac{r^2}{b^2}\right)dt^2+\left(1-\frac{r_0^n}{r^n}-\frac{m}{n+2}\frac{r^2}{b^2}\right)^{-1}dr^2
+r^2d\Omega_{n+1}^2+b^2d\Omega_{m+1}^2,
\end{equation}
solves Einstein equations
\begin{equation}
G_{\mu\nu}=\lambda g_{\mu\nu}.
\end{equation}
Here $m, n$ are two positive integers and the values of the scalar
curvature $R$ and cosmological constant $\lambda$ are related to
$m, n$ as follows
\begin{equation}
R=-\frac{m\left(m+n+4\right)}{b^2},\ \ \ \ \ \ \ \
\lambda=\frac{m\left(m+n+2\right)}{2b^2}.
\end{equation}
$b$ is a constant which has the meaning of the scale of extra
dimensions. Metric Eq.(24) describes an $(n+3)$ dimensional
Schwarzschild black hole solution which has another $(m+1)$ extra
dimensions.\\
\hspace*{7.5mm}Eq.(26) tells us the cosmological constant may be
contributed by the extra dimensions of the space. It is the
monotonic increasing function of the dimensions $m, n$ and the
monotonic decreasing function of the scale of the extra space. As
an example, we consider the simplest case, $m=1$ and $n=1$. If the
scale of the extra space is the order of Plank length $l_p$, then
the energy density contributed by the cosmological constant is
\begin{equation}
\rho_\lambda=\frac{2}{8\pi l_p^2}=4.06\times
10^{95}\left(kg/{m^3}\right),
\end{equation}
which is the same as the result of vacuum energy in the quantum
field theory [7]. It is a much
larger energy density which puzzles us for years. \\
\hspace*{7.5mm}On the other hand, if the scale of the extra space
is the order of present universe $b=1.37\cdot10^{10}l.y.$, then
the energy density contributed by the cosmological constant is
\begin{equation}
\rho_\lambda=\frac{2}{8\pi b^2}=6.37\times
10^{-27}\left(kg/{m^3}\right),
\end{equation}
approximately $2/3$ of the critical energy density of the
universe. It is the same as the result of dark energy in the
current observations [8].\\
\hspace*{7.5mm}To sum up, if we assume that the scale of extra
dimensions is the order of Plank length, then the cosmological
constant is just the vacuum energy density, which is the same as
the result of quantum field theory. On the other hand, if we
assume that the scale of extra dimensions is that of the present
universe, then the cosmological constant is just the dark energy
density, which is the same as the result of current observations.
Further more, the dimension analysis of the cosmological constant
tells us the dimension of $\lambda$ is the inverse of the square
of length. So $\lambda$ should physically originates from the
scale of some physical object. We believe this is the truth. Thus
the idea of the cosmological constant rooting in the extra
dimensions of space has some plausibility.\\
\hspace*{7.5mm}In string theory, people assume  that
gravity-together with the electromagnetic, strong and weak gauge
forces-lives everywhere in $11$-D space-time. The extra
dimesntions may be very tiny or very large. If the scale of the
extra dimensions is very large, why not we find it? A new
possibility was brought to light in 1998 by Arkani-Hamed,
Dimopoulos and Dvali [9]. They think that the electromagnetic,
weak and strong forces, as well as all the matter in the universe,
would be trapped on a surface with our three spatial dimensions,
like dust particles on soap bubbles. Only gravitons would be able
to leave the surface and move throughout the full volume. This 3-D
surface is known as a "brane". Since our everyday experiences are
prejudiced by electromagnetism, which is trapped on the brane. We
naturally do not see extra dimensions in everyday life. Meanwhile,
the highest energy particle accelerators extend our range of sight
to include the weak and strong forces down to small scales, around
10-15 mm. We may therefore be blissfully unaware of any extra
dimensions. \\
\hspace*{7.5mm}The only force we can use to probe gravity-only
extra dimensions is, of course, gravity itself. Unfortunately,  we
have almost no knowledge of gravity at distances less than about a
millimetre. This is because the direct tests of the gravitational
force are based on torsion-balance experiments that measure the
attraction between oscillating spheres. The smallest scale on
which this type of tabletop experiment has so far been performed
is 0.2 mm. Hence, only when much smaller than 1 mm, we are able to
detect the distinct effects of gravitons propagating into the
extra dimensions.
\section{conclusion and discussion} \hspace*{7.5mm}In conclusion,
we have extended the McVittie solution from four dimensions to
arbitrary dimensions i.e. Eq.(17). It should be note that we make
an important step which makes the deriving very simple, i.e.
Eq.(12). When the mass of the black hole is set to zero, the
metric recovers to the FRW metric with arbitrary dimensions. On
the hand, when the scale factor $a(t)$ is set to a constant and
the curvature of the spacetime zero, the metric recovers to the
higher dimensional Schwarzschild metric. Thus the solution
describes the higher dimensional Schwarzschild black hole which is
embedded in the FRW
universe.\\
\hspace*{7.5mm}For simplicity in mathematics, we only rewrite the
higher dimensional Schwarzschild-de Sitter metric from the cosmic
coordinates system to the Schwarzschild coordinates system. As an
even general extending, we find an even more generalized
Schwarzschild-de Sitter metric which is affixed another extra
dimensions. Our solution shows that the extra dimensions might
play the role of the cosmological constant or dark energy. This
supports the hypothesis which is proposed by Gu and Hwuang.\\
\hspace*{7.5mm}After this paper was published, Yoserf Verbin
informed me that the solution Eq.(17) is actually the Patel-Tikekar-Dadhich solution [10].\\
 \hspace*{7.5mm}I thank Dr. Verbin for bring this Ref.[10] to my attention.
\ \ \ \ \ \ \\
\hspace*{7.5mm}\acknowledgements I am very grateful to the two
anonymous referees. One of them pointed out one mistake in my
original manuscript. Their expert and insightful comments, which
have certainly improved the paper significantly. This study is
supported in part by the Special Funds for Major State Basic
Research Projects and by the National Natural Science Foundation
of China and the Postdoctoral Science Foundation of China.

\end{document}